\documentclass[12pt,a4paper]{article}
\usepackage{amsbsy,amssymb,graphicx}
\renewcommand{\baselinestretch}{1.8}

\begin{document}
\def\alt{\lesssim}
\def\agt {\gtrsim}
\thispagestyle{empty}
\begin{flushright} SNUTP 00-007 \\ 
\end{flushright}

\def\baselinestretch{1.2}
\centerline{\Large \bf On mass dependences of the one-loop effective action} 
\centerline{\Large \bf in simple backgrounds}
\vskip 1cm 
\centerline{O-Kab Kwon\footnote{Electronic address; kok@phya.snu.ac.kr}
 and Choonkyu Lee\footnote{Electronic address; cklee@phya.snu.ac.kr}} 
\begin{center}
{\it Physics Department and Center for Theoretical Physics, 
Seoul National University \\ Seoul 151-742, Korea}
\end{center}
\centerline{Hyunsoo Min\footnote{Electronic address; hsmin@dirac.uos.ac.kr}}
\begin{center}
{\it Physics Department, University of Seoul, Seoul 130-743, Korea}
\end{center}
\vskip 3cm
\begin{abstract}
If background fields are soft on the scale set
by mass of the particle involved, a reliable approximation 
to the field-theoretic one-loop effective action is obtained by 
a systematic large mass expansion involving higher-order 
Seeley-DeWitt coefficients. Moreover, if
the small mass limit of the effective action in a
particular background has been found by some other 
means, the two informations may be used to infer the corresponding 
result for {\em arbitrary} mass values. This
method is used to estimate the one-loop contribution 
to the QCD vacuum tunneling amplitude by quarks of 
arbitrary mass.
\end{abstract}
\newpage

Quantum or loop corrections to the effective action,
in general or specific background fields, are of fundamental
importance in field-theoretic studies of many physical
processes. The most well-known example is the exact one-loop 
QED effective action for electrons in a uniform electromagnetic field background,
computed first by Euler and Heisenberg\cite{Euler} and many others 
since \cite{Schwinger,Nikisnov}: 
this provides us with valuable information on the vacuum
polarization phenomenon and on the electron-positron pair production from 
the vacuum. 
Also interesting  physical effects have been demonstrated by studying the 
one-loop correction to the effective action in a soliton or instanton
background \cite{Goldstone,tHooft76}.

In four-dimensional field theory contexts, however, the {\em
exact} computation of the 
one-loop effective action in any non-trivial 
background field generally corresponds
to a formidable mathematical problem. A well-known 
approximation scheme in this regard is the so-called
derivative expansion\cite{Nikisnov,Chan} of the effective action, which 
may be used for a sufficiently smooth background 
field. In this paper, we discuss the possibility of utilizing 
a large mass expansion (for which simple computer algorithms have been 
developed recently) and mass interpolation to find the one-loop
effective action for an arbitrary mass parameter. Euclidean four-dimensional
space-time is assumed below.

To explain our approach, consider the one-loop effective action
$\Gamma(A)$   for a complex spin-0 field of mass $m$ in some 
Yang-Mills background fields $A_\mu^a(x)$. The quadratic differential
operator appropriate to the scalar field is
\begin{equation}
G^{-1}+m^2  =  
-D_\mu D_\mu +m^2\quad
(\equiv -D^2 +m^2)
\end{equation}
(with  $D_\mu=\partial_\mu -i A^a_\mu T^a \equiv \partial_\mu-iA_\mu$), and
the corresponding background-free one is $G^{-1}_0 +m^2=-\partial^2 + m^2$.
The Pauli-Villars regularized form of the effective action can then be 
expressed
as
\begin{eqnarray}
\Gamma(A) &=&  \ln \left[
{{\rm Det}(G^{-1} + m^2) \over {\rm Det} (G^{-1}_0+m^2)}
{{\rm Det}(G^{-1}_0 + \Lambda^2) \over {\rm Det} (G^{-1}+\Lambda^2)}
\right]  \nonumber \\
&=& - \int^\infty_0 { ds \over s} 
(e^{-m^2 s} - e^{-\Lambda^2 s} ) \int d^4x \, {\rm tr}
 \left[<xs|x> - <xs|x>|_{A_\mu=0} \right]. \label{effectiveaction}
\end{eqnarray}
Here the second expression is the Schwinger proper-time 
representation\cite{Schwinger} which 
involves the coincidence limit of the proper-time Green function,
$<x s|y> \equiv <x| e^{-s G^{-1}}|y> $. The latter
admits the small-$s$ asymptotic expansion of the form \cite{DeWitt,Seeley}
\begin{equation}\label{series}
s\to0+: \qquad <xs|y>={1 \over (4\pi s)^2} e^{-{(x-y)^2 \over 4s}}
\left\{ \sum^\infty_{n=0} s^n a_n(x,y)\right\},
\end{equation}
with $a_0(x,x)=1$.

Using the expansion (\ref{series}) in (\ref{effectiveaction}), one finds that 
the divergent terms of $\Gamma(A)$ as $\Lambda\to \infty$ are 
related to the first and second Seeley-DeWitt coefficients,
$\tilde a_1(x)\equiv {\rm tr} a_1(x,x)$ and 
$\tilde a_2(x)\equiv {\rm tr} a_2(x,x)$. Simple calculations yield
\begin{equation}
\tilde a_1(x)=0,~~~~ 
\tilde a_2(x)= -{1\over12}\, {\rm tr}
 ( F_{\mu\nu}(x) F_{\mu\nu}(x)), 
\end{equation} where 
 $F_{\mu\nu}\equiv F_{\mu\nu}^a T^a=
 i[D_\mu,D_\nu]$. Hence the above effective action can be cast as
\begin{equation}\label{gamma}
\Gamma(A)={1\over12}{C\over (4\pi)^2}
(\ln{\Lambda^2 \over m^2}) \int
d^4x F^a_{\mu\nu} F^a_{\mu\nu}  + \overline{\Gamma}(A),
\end{equation}
($C$ is defined by ${\rm tr}(T^aT^b)=\delta_{ab} C$), where the amplitude
\begin{equation}\label{gammabar}
\overline{\Gamma}(A)=- 
\int^\infty_0 {ds \over s^3} e^{-m^2s}\int d^4x 
\left[1-\left.\left(1+s{\partial \over \partial s}+ {1\over2} 
s^2{\partial^2 \over \partial s^2}\right)\right|_{s=0}\right]
{\rm tr} \left(s^2 <xs|x>\right)
\end{equation}
is well-defined as long as $m^2\ne0$. [In (\ref{gammabar}),
$(1+s{\partial \over \partial s}+ {1\over2} 
s^2{\partial^2 \over \partial s^2})|_{s=0} f(s)\equiv
f(0)+sf'(0) + {1\over2} s^2 f''(0)$].
The logarithmic divergence in (\ref{gamma}) is canceled by the
renormalization  counterterm associated with the coupling constant 
renormalization of the classical action
${1\over4g_0^2} \int d^4x
F^a_{\mu\nu}F^a_{\mu\nu} $. But the result is renormalization-prescription 
dependent. In fact, our amplitude 
$\overline{\Gamma}(A)$ in (\ref{gammabar}) can be viewed  as {\em a}  
renormalized one-loop effective action for $m^2\neq 0$.
If one instead adds to $\Gamma(A)$ the counterterm
$\Delta \Gamma(A)=-{1\over12}{C\over (4\pi)^2}(\ln{\Lambda^2\over \mu^2}) 
\int d^4x   F^a_{\mu\nu}F^a_{\mu\nu}$
($\mu$ is an arbitrarily introduced renormalization mass),
the resulting
renormalized one-loop effective action reads 
\begin{equation}\label{gammaren}
\Gamma_{\rm ren}(A)=-{1\over12}{C\over (4\pi)^2}(\ln{m^2\over \mu^2}) \int d^4x   
F^a_{\mu\nu}F^a_{\mu\nu} + \overline{\Gamma}(A),
\end{equation}
which has now a well-defined limit even for $m^2\to 0$. For the expression  
 in the minimal subtraction\cite{MS} in the 
dimensional regularization scheme, a further finite renormalization counterterm
should be introduced\cite{tHooft76}. These  
differences in the renormalized expressions reflect 
different ways of defining the renormalized coupling.

The next task will be to find the full finite amplitude for the one-loop
effective action; for any non-trivial background field, this is very 
difficult. If the mass $m$ is relatively
large, however, a large-mass expansion obtained by inserting the
asymptotic series (\ref{series}) into (\ref{gammabar}) can be useful:
\begin{equation} \label{gammabarmass}
\overline{\Gamma}(A)=-{1\over (4\pi)^2} \sum_{n=3}^\infty
{(n-3)! \over (m^2)^{n-2} }\int d^4 x \tilde{a}_n(x), \hskip 1cm 
(\tilde{a}_n(x)\equiv {\rm tr} a_n(x,x) ).
\end{equation}  
For $\Gamma_{\rm ren}(A)$, one may use
the formula (\ref{gammaren}) together with this
expansion.
Thus, for relatively large mass, the one-loop effective action can be 
approximated by a series involving higher-order Seeley-DeWitt coefficients
$\tilde a_n(x)$ ($n\ge3$), for which 
computer algorithms are now available\cite{Fliegner}.
The useful range of this expansion, as regards the magnitude of $m$, will
depend much on the nature of the background field and
on the characteristic scale entering the background.

In this work we are interested in the one-loop
effective action in some physically important background as a
{\em function of mass parameter} $m$. Even for a simple  
background field, it will not be possible to infer the complete $m$-dependence 
on the basis of the series(\ref{gammabarmass}) alone; the series loses
the predictive power for `small' values of $m$. Actually, as we shall see 
below, this large mass expansion (truncated at certain order) appears to provide a 
surprisingly good approximation even for only moderately large values of $m$. 
Then, if the effective action in the small mass limit became known by
independent methods (possibly exploiting certain symmetry present in the zero 
mass case), one may hope that a reliable interpolation between the small-mass
and relatively large-mass expressions could be made to obtain 
a reasonable fit over 
the entire mass values.
Below, we shall first test this idea with the constant Yang-Mills field strength 
background case for which the exact one-loop effective action
is known. The same method will then be applied to the case of 
significant interest---we estimate the one-loop instanton contribution
to the QCD vacuum tunneling \cite{tHooft76,CDG} by quarks of arbitrary mass.
The QCD vacuum tunneling amplitude due to quarks
of vanishingly small mass was calculated analytically by 
'tHooft\cite{tHooft76}; this result may be relevant to $u$- and $d$-quarks,
but not for others.

In the case of non-Abelian gauge theories, a constant field strength
can be realized  either by an Abelian vector potential which varies linearly
with $x^\mu$  or by a constant vector
potential whose components do not commute\cite{Leutwyler}.
In this paper we only consider the
case of the Abelian vector potential.
Assuming SU(2) gauge group, an Abelian vector potential can then 
be written as
$A_\mu=-{1\over4} f_{\mu\nu}x_\nu \tau_3$ 
(with the field strength tensor
$F_{\mu\nu}=f_{\mu\nu} \tau_3/2$), where 
$\tau^3$ is the third Pauli matrix. 
If we further restrict our attention to the self-dual  
case, we can set  $f_{23}=f_{41}=H$ with the constant `magnetic' field $H$.

In this Abelian constant
self-dual field, let us consider the one-loop effective action 
induced by isospin-1/2, spin-0 matter fields, taking the mass $m$ of
our spin-0 fields to be relatively large so that
the large mass expansion (\ref{gammabarmass}) may be used. 
From the result of 
\cite{Fliegner}, 
some leading  Seeley-DeWitt coefficients are easily evaluated for this 
case:
\begin{equation}
\tilde{a}_4(x)={2\over 15} (H/2)^4,  \quad
\tilde{a}_6(x)=-{4\over 189} (H/2)^6, \quad
\tilde{a}_8(x)={2\over 675}(H/2)^8
\end{equation}
[Note that we get zero for all odd coefficients here].
Using these values, we then find for relatively large $m$ the 
expression
\begin{equation}\label{inversemass}
\overline{\Gamma}(H;m)= - {VH^2 \over 8\pi^2}
\left( {1\over240} ({H\over m^2})^2
- {1\over1008 } ({H\over m^2})^4
+{1\over1440} ({H\over m^2})^6
+\cdots \right),
\end{equation}
where $V$ denotes the four-dimensional Euclidean volume. 
For this case, it is actually not difficult to find the  {\em exact}
expression for the one-loop
effective action, following rather closely  Schwinger's original analysis in
QED\cite{Schwinger}: the result for $\overline{\Gamma}$ 
(see (\ref{gammabar})) turns out to be
\begin{equation} \label{constbar}
\overline{\Gamma}(H;m)=- 2V \int^\infty_0 {ds\over s}
 e^{-m^2 s} {1\over(4\pi s)^2}
\left[{(Hs/2)^2 \over \sinh^2(Hs/2)} -1+ {1\over3} (Hs/2)^2 \right].
\end{equation}
Comparing the result of large mass expansion in (\ref{inversemass})
against this exact expression, we can investigate the validity 
range of the former. From the plots in Fig.1, it should be
evident that for mass value in the range $m/\sqrt{H}\agt 1$,
summing only a few leading terms in the series (\ref{inversemass}) already
produces values which are very close to the exact ones. 
\begin{figure}
\begin{center}
\includegraphics[height=3in]{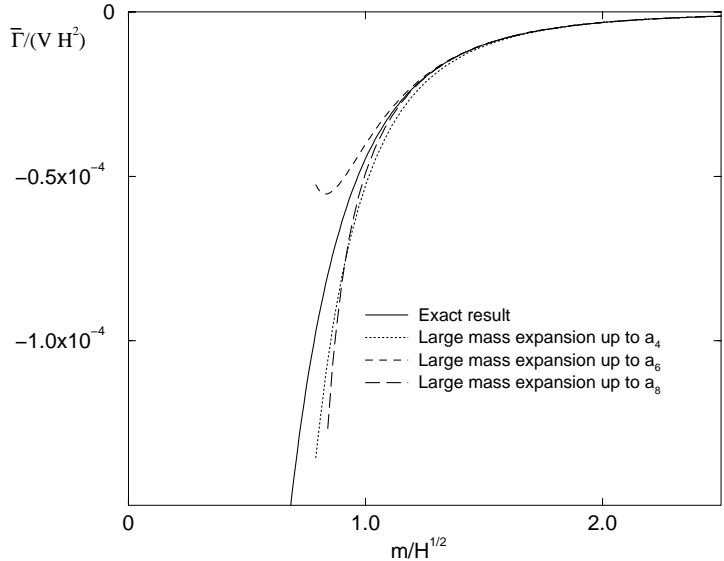}
\caption{ Plot of the effective action $ \overline\Gamma(H;m)$.}
\end{center}
\end{figure}

Now suppose that the exact expression (\ref{constbar}) were not 
available to us. For mass value which is not so large
(i.e., if $m/\sqrt{H}<1$), large mass expansion (\ref{inversemass}) fails
to give useful information. Nevertheless, if one happens to know the
one-loop effective action for {\em small} mass, this additional
information and the large mass expansion may well be used to infer
the behavior of the effective action for 
general, small or large, mass. In exhibiting this, 
$\overline{\Gamma}(H;m)$ will not be convenient since it becomes
ill-defined as $m\to0$. So, based on the relation (\ref{gammaren}), we may 
consider the renormalized action $\Gamma_{\rm ren}(H;m,\mu)$ given by 
\begin{equation}\label{constren}
\Gamma_{\rm ren}(H;m,\mu)=-  { V H^2 \over(4\pi )^2 \cdot 6}
\ln({m^2 \over \mu^2})
 + \overline{\Gamma}(H;m).
\end{equation}
which is well-behaved for small $m$. Large mass expansion for 
$\Gamma_{\rm ren}(H;m,\mu)$ results once if the expansion (\ref{inversemass})
is substituted in the right hand side of (\ref{constren}). On the other hand,
$\Gamma_{\rm ren}(H;m,\mu)$ has the small-$m$ expansion 
(which is extracted using (\ref{constbar})),
\begin{equation}\label{gammarenzero}
\Gamma_{\rm ren}(H;m=0,\mu)=
VH^2\left({1\over(4\pi)\cdot 6}\ln{\mu^2\over H} +0.00209
 - 0.00633({m^2\over H}) + \ldots \right).
\end{equation}
In combining these two informations from different mass ranges, it is 
convenient to  consider the $\mu$-independent quantity (especially 
for mass interpolation purpose)
\begin{equation}\label{gammatilde}
\tilde{\Gamma}(H;m)\equiv \Gamma_{\rm ren}(H;m,\mu)
-\Gamma_{\rm ren}(H;m=0,\mu).
\end{equation}
In Fig.2, the graph for $\tilde{\Gamma}(H;m)$
has been given as a function of
$m/\sqrt{H}$. The exact result, represented by a solid line, 
exhibits a monotonically decreasing behavior starting from the maximum at 
$m/\sqrt{H}=0$. As we mentioned already, the large mass expansion can be 
trusted in the range $m/\sqrt{H}\agt 1$. This curve may then be smoothly 
connected to that given from the small-$m$ expansion (\ref{gammarenzero}),
assuming a monotonic behavior (as should be reasonable for a simple 
background field). Evidently, with this interpolation, one could have 
acquired a nice overall fit over the entire mass range if the exact 
curve were not known.

\begin{figure}
\begin{center}
\includegraphics[height=3in]{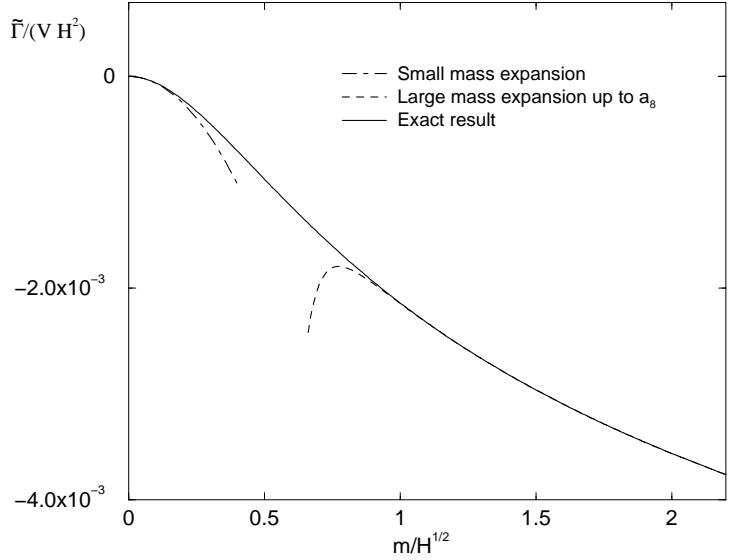}
\caption{Plot of  $\tilde\Gamma (H;m)$. }
\end{center}
\end{figure}

%
%
%

Now turn to the case of a BPST instanton background\cite{Belavin},
i.e., a self-dual solution of Yang-Mills field equations given by 
\begin{equation}\label{instansol}
A_\mu(x)\equiv A^a_\mu(x){\tau^a\over 2}= { \eta^{(+)}_{\mu\nu a}\tau_a 
x_\nu \over
x^2 +\rho^2},
\end{equation}
where $\eta^{(+)}_{\mu\nu a}\;(a=1,2,3)$ are the so-called 'tHooft 
symbols\cite{tHooft76}. With QCD in mind, the effective action due to a 
spin-1/2 quark field (in the fundamental representation) with unspecified 
mass $m$ will be of special interest. Here we define the proper-time Green 
function by $<xs|y>= <x|e^{-s(\gamma_\mu D_\mu)^2}|y>$ (our antihermitian 
$\gamma$-matrices satisfy the relations $\{\gamma_\mu,\gamma_\nu\} 
= -2\delta_{\mu\nu}$), so that we may have the spin-1/2 one-loop 
effective action expressed as 
\begin{equation}\label{spinoreff}
\Gamma^{(1/2)}(A)={1\over 2}  \int^\infty_0 { ds \over s} 
(e^{-m^2 s} - e^{-\Lambda^2 s} ) \int d^4x \, {\rm tr}
 \left[<xs|x> - <xs|x>|_{A_\mu=0} \right]. \label{effectiveactionhalf}
\end{equation}
For the corresponding $\tilde{a}_2$-coefficient, we have $\tilde{a}_2 
= {2\over 3} {\rm tr}(F_{\mu\nu}(x)F_{\mu\nu}(x))$. 
So the renormalized one-loop 
effective action $\Gamma^{(1/2)}_{\rm ren}(A)$---the direct spin-1/2 
analogue of (\ref{gammaren})---can be obtained if the counterterm 
$\Delta\Gamma(A) = - {1\over 3}{C\over (4\pi)^2}(\ln{\Lambda^2 \over 
\mu^2})\int d^4xF^a_{\mu\nu}F^a_{\mu\nu}$ is added to the unrenormalized 
expression (\ref{spinoreff}). Note that if the Dirac operator 
$\gamma_\mu D_\mu$ possesses normalizable zero modes \cite{zeromode},
the renormalized quantity $\Gamma^{(1/2)}_{\rm ren}(A)$ is still infrared 
divergent at $m^2 = 0$. Actually, based on the hidden supresymmetry present 
in a self-dual Yang-Mills background\cite{tHooft76}, it is possible to 
derive a following simple relationship\cite{selfdualym} existing between 
the spin-1/2 and spin-0 one-loop effective actions:
\begin{equation}
\Gamma^{(1/2)}_{\rm ren}(A) = -{1\over2} n_F 
\left(\ln{m^2\over\mu^2}\right) 
-2 \Gamma_{\rm ren}(A),
\end{equation} 
or, for the respective contributions to the tunneling amplitude, 
\begin{equation}
e^{-\Gamma^{(1/2)}_{\rm ren}(A)} = \left({m \over \mu}\right)^{n_F} 
e^{2\Gamma_{\rm ren}(A)}.
\end{equation}
Here, $n_F$ is the number of normalizable spinor zero modes in the given 
self-dual Yang-Mills background, and $\Gamma_{\rm ren}(A)$ the corresponding 
one-loop effective action(defined in accordance with (\ref{gammaren}) above)
for a `spin-0 quark' of the same mass $m$. Due to this relationship, our 
problem is again reduced to that of a spin-0 field.(To obtain the spin-1/2 
one-loop effective action in the minimal subtraction in the dimensional 
regularization scheme, the finite renormalization counterterm 
$\Delta \Gamma(A)^\prime ={C\over(4\pi)^2 \cdot 3} (\ln4\pi -\gamma)
\int d^4x  F^a_{\mu\nu}F^a_{\mu\nu}$ 
($\gamma=0.5772\ldots$ is the Euler's constant) must be added further
to that of $\Gamma_{\rm ren}^{(1/2)}(A)$\cite{tHooft76}.)

For relatively large mass $m$, the renormalized effective action for a 
spin-0 matter field in the instanton background (\ref{instansol}) can be 
studied with the help of large-mass asymptotic series (\ref{gammabarmass}) 
for $\overline\Gamma$. Note that $\overline\Gamma$ is a function of 
$m\rho$ only. The
coefficient $a_3(x,x)$ is 
\begin{equation}
a_3(x,x)={1\over 120}D_\mu F_{\nu\lambda} D_\mu F_{\nu\lambda}
-{i\over45}F_{\mu\nu}F_{\nu\lambda}F_{\lambda\mu},
\end{equation}
and so, for the instanton background (\ref{instansol}), we obtain the 
result  
$\int d^4x \;\tilde{a}_3(x) = {16\over 75} {\pi^2\over \rho^2}$. Calculations 
of higher-order Seeley-DeWitt coefficients with the instanton background 
can be very laborious.(For $a_6(x,x)$ for instance, the full expression 
occupies more than a page\cite{Fliegner}). Together with the formulas given 
in ref.\cite{Fliegner}, we have thus used the 
"Mathematica" program to do the necessary trace calculations and also tensor 
algebra. The results are as follows:
\begin{eqnarray}
\int d^4x\;\tilde{a}_4(x) 
= {272\over 735}{\pi^2\over \rho^4},
\qquad\int d^4x\,\tilde{a}_5(x)=-{1856\over2835} {\pi^2\over\rho^6},
\qquad \int d^4x\,\tilde{a}_6(x)={63328\over444675} {\pi^2\over\rho^8}.
\end{eqnarray}
From these we obtain the following expression for $\bar{\Gamma}(m\rho)$:
\begin{equation} \label{gammabarlargemass}
\overline{\Gamma}(m\rho)=-{1\over75}{1\over m^2\rho^2}
-{17\over735}{1\over m^4\rho^4}
+{232\over2835}{1\over m^6\rho^6}
-{7916\over148225}{1\over m^8\rho^8}
+\cdots
\end{equation}
\begin{figure}
\begin{center}
\includegraphics[height=3in]{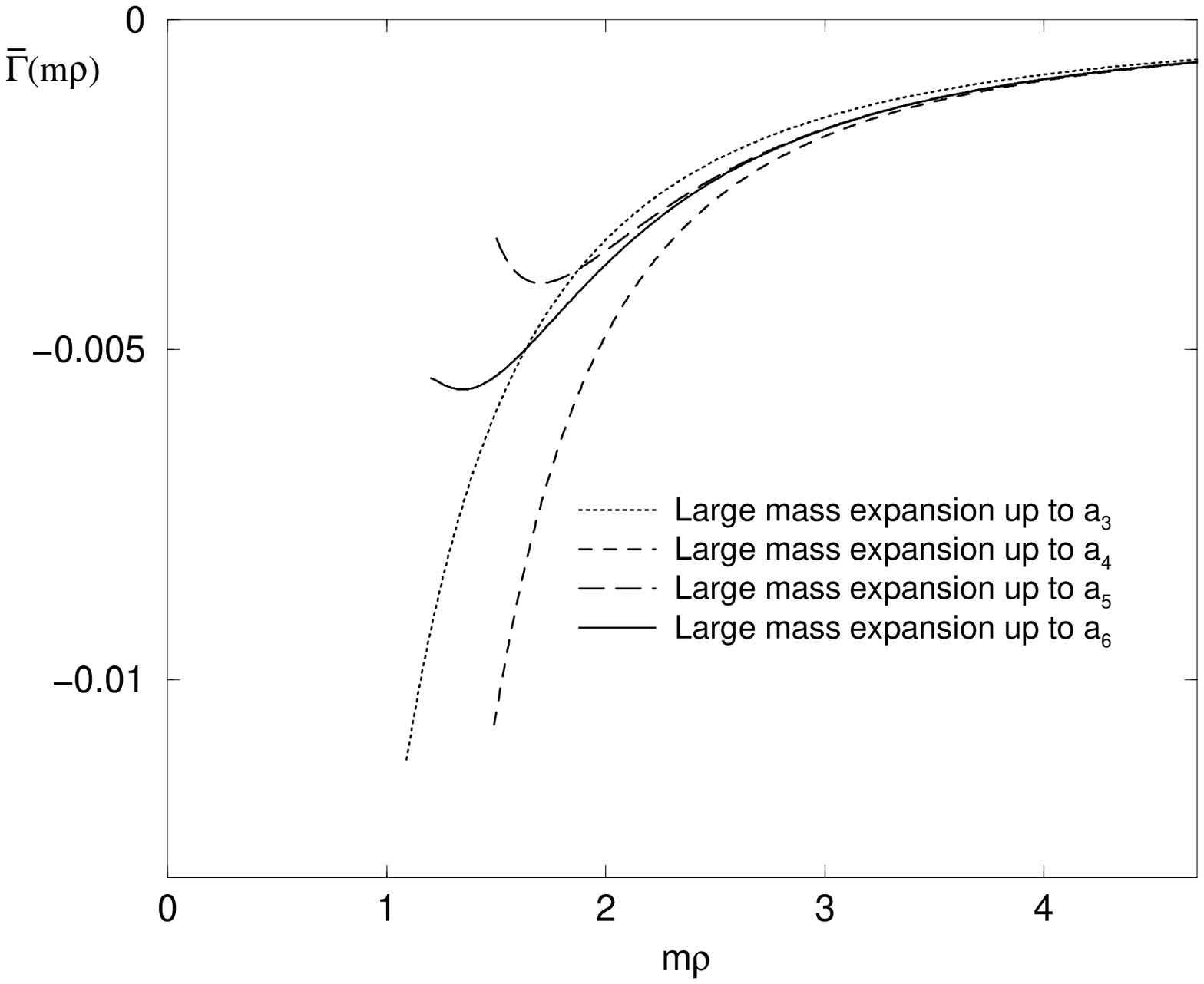}
\caption{Plot of $\overline{\Gamma}(m\rho)$ for the instanton background.}
\end{center}
\end{figure}
Plotting this expression (first keeping only the $a_3$-term, then including 
the $a_4$-term also, etc), we find that the curve is quite stable if 
$m\rho \agt 1.5$. (See Fig.3). The 
result of large mass expansion may thus be trusted in the mass range given by 
$m\rho\agt 1.5$.

To compare the above findings with the small-mass expression, we again consider 
the renormalized effective action $\Gamma_{\rm ren}(A)$, which is denoted in 
the instanton background (\ref{instansol}) by $\Gamma_{\rm ren}(m,\rho,\mu)$. From
 (\ref{gammaren}), we have 
\begin{equation}
\Gamma_{\rm ren}(m,\rho,\mu) = -{1\over6} \ln {m\over\mu} 
  + \overline{\Gamma}(m\rho).
\end{equation}
Then,
from the computations of 'tHooft\cite{tHooft76} and of ref.\cite{Carlitz},
$\Gamma_{\rm ren}(m,\rho,\mu)$ for sufficiently small values of $m\rho$ is 
approximated by 
\begin{equation}\label{gammarenins}
\Gamma_{\rm ren}(m,\rho,\mu) = {1\over 6}\ln \mu\rho + 
\alpha({1\over2})+{1\over 2}(m\rho)^2 \ln m\rho + \cdots
\end{equation}
with $\alpha({1\over2}) =  {1\over6} \gamma +
{1\over6}\ln \pi - {1\over \pi^2}\zeta^\prime(2) -{17\over72}$, where 
$\zeta^\prime(s)$ is the first derivative of Riemann zeta function. We also 
define the $\mu$-independent quantity
\begin{eqnarray}\nonumber 
\tilde{\Gamma}(m\rho) &\equiv& \Gamma_{{\rm ren}}(m,\rho,\mu) -
\Gamma_{{\rm ren}}(0,\rho,\mu)\\
 &=& -{1\over6} \ln {m\rho} - \alpha({1\over2}) +
 \overline{\Gamma}(m\rho).
\label{gammatilda}
\end{eqnarray}
For sufficiently small $m\rho$, we have $\tilde{\Gamma}(m\rho) \simeq 
{1\over 2}(m\rho)^2\ln m\rho$ ; but, for $m\rho \agt 1.5$, a good 
approximation to $\tilde{\Gamma}(m\rho)$ results if (\ref{gammabarlargemass}) 
is used in the second form of (\ref{gammatilda}). These small-mass and relatively 
large-mass expressions for  $\tilde{\Gamma}(m\rho)$ are plotted in Fig.4. 
\begin{figure}
\begin{center}
\includegraphics[height=3in]{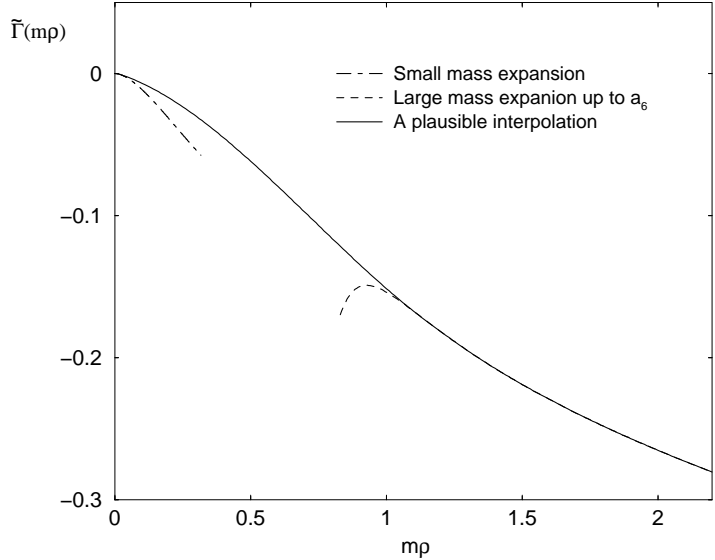}
\caption{Plot of $\tilde{\Gamma}(m\rho)$ for the instanton background.}
\end{center}
\end{figure}
Also 
included is the smooth interpolating curve connecting the two regions, assuming the 
monotonousness in the range $0< m\rho \alt 1.5$. In view of a simple character of 
the background field (\ref{instansol}) and the fact that $m\rho$ is the sole 
relevant variable for $\tilde{\Gamma}$, we believe that the latter 
assumption is very plausible.  For further support on this, we need an 
improved approximation(i.e., beyond (\ref{gammarenins})) for small mass;
this is left for future investigation. Besed on this interpolation, 
one might also go on to devise a simple functional form for 
$\Gamma_{\rm ren}(m,\rho,\mu)$(approximately valid for {\em any} mass value) 
for phenomenological studies concerning instanton effects.

\vskip .4in
{\bf Acknowledgements}
\vglue .2in

This work was supported in part by the BK 21 project of the Ministry of 
Education, Korea(O.K. and C.L.) and by KOSEF Grant 97-07-02-02-01-3(C.L.). 



\begin{thebibliography}{99}

\bibitem{Euler}W.Heisenberg and H.Euler, Z. Phyzik {\bf 98}, 714 (1936). 
\bibitem{Schwinger}J. Schwinger, Phys. Rev. {\bf 82}, 664 (1951).
\bibitem{Nikisnov}A. Nikisnov, Sov. Phys. JETP {\bf 30}, 660 (1970); 
W. Ditrich, J. Phys. {\bf A9},   (1976);
W. Greiner and J. Reinhardt, {\it Quantum Electrodynamics} (Springer, 1992).
\bibitem{Goldstone}J.Goldstone and F.Wilczek, Phys. Rev. Lett. {\bf 47}, 986 (1981);
I.J.R. Aitchison and C.M.Fraser, Phys. Rev. {\bf D31}, 2605 (1985).
\bibitem{tHooft76}G.'tHooft, Phys. Rev. {\bf D14}, 3432 (1976);
(E) {\bf D18}, 2199 (1978).  
\bibitem{Chan}L.-H.Chan, Phys. Rev. Lett. {\bf 54}, 1222 (1985);
H.W.Lee, P.Y.Pac and H.K.Shin, Phys. Rev. {\bf D40}, 4202 (1989);
D.G.C.McKeon, Ann. Phys. {\bf 224}, 139 (1993);
V.P.Gusynin and I.A.Sovkovy, Can. J. Phys. {\bf 74}, 282 (1996);
L.-H.Chan, Phys. Rev. {\bf D55}, 6223 (1997);
M.Reuter, M.G.Schmidt and C.Schubert, Ann.Phys. {\bf 259}, 313 (1977);
G.V.Dunne, ``Derivative Expansion and Soliton Mass'', hep-th/9907208. 
\bibitem{DeWitt}B.S.DeWitt, {\it Dynamical Theory of Groups and Fields} 
(Gordon and Breach, New York, 1965);
Phys. Rep. {\bf 19}, 295  (1975); J. Honerkamp, Nucl. Phys. {\bf B36}, 130 (1970).
\bibitem{Seeley}R.T.Seeley, Proc. Symp. Pure Math. {\bf 10}, 288 (1967);
P.B.Gilkey, Invariance Theory, the Heat Equation and the Atiyah-Singer Index 
Theorem(Wilmington, DE, Publish or Perish, 1984).
\bibitem{MS} G.'tHooft and M.Veltman, Nucl. Phys. {\bf 44}, 189 (1972).
\bibitem{Fliegner}
A.A. Bel'kov, D. Ebert, A.V. Lanyov and  A.Schaale, Int. J. Mod. Phys. {\bf C4},
 775 (1993);
A.A. Bel'kov, A.V. Lanyov and  A.Schaale, Comput. Phys. Comm. {\bf95}, 123 (1996);
D.Fliegner, M.G.Schmidt and C.Schubert, Z. Phys. {\bf C64},
111 (1994); 
D.Fliegner, P.Haberl, M.G.Schmidt and C.Schubert, Ann. Phys. {\bf 264}, 51 (1998);
M.J.Booth, hep-th/9803113.
\bibitem{CDG} C.G. Callan, R. Dashen and D.J. Gross, Phys. Lett.
{\bf 63B}, 334 (1976);
R. Jackiw and C. Rebbi, Phys. Rev. Lett. {\bf 37}, 172 (1976).
\bibitem{Leutwyler}L.S.Brown and W.I.Weisberger, Nucl. Phys. {\bf B157},
285 (1979);
H.Leutweyler, Phys. Lett. {\bf 96B},154 (1980); Nucl. Phys.
{\bf B179}, 129 (1981).
\bibitem{Belavin}A.Belavin, A.Polyakov, A.Schwartz and Y.Tyupkin, Phys. Lett. {\bf 59B},
85 (1975).
\bibitem{zeromode}J.Kiskis, Phys. Rev.{\bf D15}, 2329 (1977); L.S.Brown, 
R.D.Carlitz and C.Lee, Phys. Rev. {\bf D16}, 417 (1977).
\bibitem{selfdualym}C.Lee, H.W.Lee, and P.Y.Pac, Nucl. Phys.{\bf B201}, 429 (1982). 
\bibitem{Carlitz}R.D. Carlitz and D.B. Creamer, Ann. Phys. (N.Y.) {\bf 118}, 429 (1979). 
\end{thebibliography}
\end{document}